\documentclass[prl,twocolumn,showpacs]{revtex4}
\usepackage{graphicx}
\begin{document}

\title{Search For the Origin of the Away-side Double-humped Structure of jets in Heavy Ion Collisions }
\author{H. Zheng, L. L. Zhu, C. B. Yang
\footnote{Email: cbyang@mail.ccnu.edu.cn}} \affiliation{ Institute
of Particle Physics, Hua-Zhong Normal University, Wuhan 430079, P.
R. China\\ Key Laboratory of Quark and Lepton Physics (CCNU),
Ministry of Education, P. R. China }

\begin{abstract}
A model is constructed for the origin of the double-humped
structure found in the di-hahron azimuthal correlation on the
away-side in heavy ion collisions. The parameters in the model are
determined by fitting $\Delta\phi$ azimuthal distribution in
central Au+Au collisions at RHIC given by STAR collaboration when
the trigger momentum is in $3-4\rm {GeV/c}$ and the associated
particle momentum is in $1.3-1.8\rm{GeV/c}$. Then we apply our
model to the semi-central and peripheral Au+Au collisions at RHIC
and the responding $\triangle \phi$ azimuthal distribution are
reproduced respectively. Our results show that the transverse flow
effect plays an essential role in the experimentally observed
di-hadron azimuthal distribution structure. The spread of the hard
parton creation point and initial moving direction is also
necessary for the double-humped structure.\pacs{25.75.Dw}
\end{abstract}

\maketitle
\section{I. Introduction}
The critical energy density predicted by Lattice Quantum
Chromodynamics (LQCD) for the formation of Quark Gluon Plasma
(QGP) \cite{fk6981992002} has been exceeded in central Au+Au
collisions at Relativistic Heavy Ion Collider (RHIC) \cite{star1}.
Some novel phenomena, such as jet-quenching and collective flows
\cite{star2, star3}, indicate the dense medium is produced in
central Au+Au collisions at RHIC \cite{star4}. A natural subject
is to study the properties of the produced medium through the
interactions of jets and the medium.

In recent years, correlations among hadrons has been one of very
active subjects, because it is a valuable tool for studying the
interactions between jets and the medium produced in heavy ion
collisions. Measurements of high transverse momentum particle
spectra provide the first striking experimental evidence of
jet-quenching. Further studies of the correlations among
intermediate $p_T$ particles reveal a complex pattern of
correlations as a function of $\Delta \eta$ and $\Delta \phi$ by
STAR Collaboration \cite{wong2009star5, wong2009star6,
wong2009star7, wong2009star8, wong2009star9, wong2009star10,
wong2009star11, wong2009star12, wong2009star13, wong2009star14,
wong2009star15, wong2009star16, wong2009star17, wong2009star18},
where $\Delta \eta$ and $\Delta \phi$ are the pseudorapidity and
azimuthal angle relative to the trigger particle, respectively.
Similar $\Delta \eta$ and $\Delta \phi$ correlations have also
been observed by the PHENIX\cite{wong2009phenix1, wong2009phenix2,
wong2009phenix3} and the PHOBOS Collaboration
\cite{wong2009phobos1}. For a brief summary of the experimental
discoveries at RHIC, see \cite{phenix5}. There are several models
for the medium response to jets, such as momentum kick model
\cite{wong}, Markovian parton scattering model \cite{charleshwa},
recombination model \cite{hwa}, Mach cone \cite{cone}, Cerenkov
gluon radiation \cite{cerenkov}, jet or shower
broadening\cite{broaden}, ect, which can explain part of phenomena
on the observed jet structure.

In heavy ion collisions, there are near-side and away-side
associate-particles for a jet. For the near-side
associate-particles, there are ridge and jet components. For the
away-side ones, there exists complex structure which depends on
the trigger momentum. In this paper, we present a simple model,
similar to the momentum kick model for the ridge formation, for
the away-side associate-particles to investigate the di-hadron
$\Delta \phi$ correlations. We only focus on the physics origin of
double-humped structure for the away-side associate-particles.

This paper is organized as follows: In section II, we give a brief
introduction of our model. We show our results in section III. The
last section is for a short conclusion.

\section{II. Our Model}
In principle, a pair of back-to-back jets can be produced at any
points in the medium. The jets lose their energy during traversing
the hot dense medium. Considering the trigger bias \cite{hy09},
the near-side jets are more likely produced near the surface of
the medium. In our model, near-side jets are assumed to emerge on
the surface of the medium for simplicity. So we will not consider
the near-side ridge and jet in this paper. The medium partons and
the jets materialize into the observed particles by assuming
parton-hadron duality. We will use the same mechanism as in
\cite{wong} for the interactions among jets and the medium
partons. Between two successive collisions between hard and soft
partons, the hard parton is assumed to move freely. The free
distance of travelling is assumed to be described by a
distribution determined with a mean free path $\lambda$. So we
have two parameters, $R$ and $\lambda$, in our model for the
characteristic lengths involved. Because our focus is searching
for the origin of double-humped structure in $\Delta\phi$ on the
away-side of jets in heavy ion collisions, the absolute values of
the lengths involved are not relevant. We normalize the medium
system radius $R=1$ and the value of mean free path $\lambda$ is
in unit of $R$. Since only the azimuthal angle of final state
particles plays a role, the absolute values of involved momenta
have no importance so that we can set the initial hard parton
momentum to be $1$ and the measure momenta of all other particles
in unit of the hard parton momentum. Then one can ask the initial
momentum direction of the away-side hard partons at the point
where the parton is created. If the pair of hard partons is
created in the collision between partons with momentum fraction
$x_1$ and $x_2$ of the nucleons, with zero transverse momentum,
the jets will be back to back when $x_1=x_2$. So generally the
produced jets can deviate from back to back. In this paper, we
assume that the near-side hard parton moves parallel to the impact
parameter, and the angle $\theta$ between the initial moving
direction of an away-side hard parton and the impact parameter  is
assumed to be given by a distribution

\begin{figure}[tbph]
\includegraphics[width=0.45\textwidth]{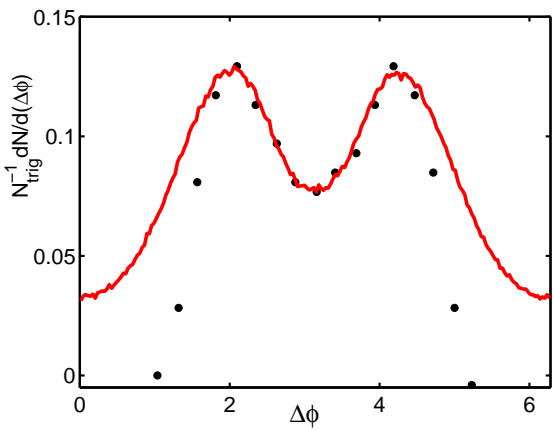}
\caption{(Color online) Double-humped structure for the away-side
associate-particle distribution. The data points (solid dots) are
from STAR collaboration \cite{star5} measured in Au+Au collisions
at 0-12\% centrality.} \label{central}
\end{figure}

\begin{equation}
f(\theta)=\frac{1}{\sqrt{2\pi
\sigma^2}}e^{-\frac{(\theta-\pi)^2}{2\sigma^2}}.\label{direction}
\end{equation}
The standard deviation is chosen to be $\sigma=\sqrt{0.5}$. When
this distribution is not used, one can study the case with exact
back to back jets in the medium. More complexity comes from a fact
that the creation point of the pair of hard partons may be
anywhere on the medium surface. In this paper, we simplify our
study by considering two cases with the hard parton's creation
point being at a fixed point or uniformly distributed on the
medium surface.

We assume that the medium partons produced in heavy ion collisions
have been thermalized, whose momentum satisfies the
two-dimensional Boltzmann distribution,
\begin{equation}
f(P)=\omega^2Pe^{-\omega P} \label{tmomentum},
\end{equation}
where $P$ is the medium parton momentum and $\omega$ is a model
parameter. The mean magnitude of momentum of the medium parton is
$\langle P\rangle=\frac{2}{\omega}$. One can sample the thermal
momentum $P_{thermal}$ of the medium parton by Eq.
(\ref{tmomentum}) with random direction. The transverse flow
effect can be considered in our model. So the medium parton may
also have a flow momentum. We assume the flow momentum is
proportional to the position vector, similar to the Hubble law
\begin{equation}
\vec {P}_{flow}=f_a\vec{r},\label{fmomentum}
\end{equation}
where $f_a$ is another model parameter. Then the momentum of a
medium parton is the sum of thermal and flow components
\begin{equation}
\vec
{P}_{parton}=\vec{P}_{thermal}+\vec{P}_{flow}.\label{pmomentum}
\end{equation}
With a mean free path $\lambda$, we assume that the stepsize $l$
between a hard parton's two successive collisions with medium
partons fulfills the distribution
\begin{equation}
f(l)=\frac{1}{\lambda}e^{-l/\lambda}.\label{mfp}
\end{equation}
After traversing $l$ the hard parton collides with a medium parton
and loses momentum, and the momentum transfer in the collision is
assumed as a simple form
\begin{equation}
\triangle \vec P=\alpha(\vec P_{jet}-\vec
P_{parton}).\label{loseenergy}
\end{equation}
After each collision, both the hard and medium partons change
their moving directions. If the hard parton momentum is smaller
than the the mean thermal momentum, the hard parton is also
assumed to materialize into the observed final state particles,
otherwise the hard parton keeps on colliding with the medium
partons until it runs out the system.

\begin{figure}[tbph]
\includegraphics[width=0.45\textwidth]{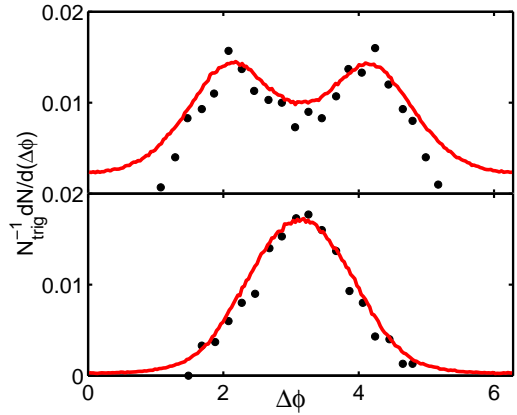}
\caption{(Color online) Azimuthal distribution of hadrons with
$2<P_T<3 \rm{GeV/c}$ associated with trigger hadrons with
$2.5<P_T<3 \rm{GeV/c}$. The data points (solid dots) are from
PHENIX collaboration \cite{phenix4} measured in AuAu collisions.
Up panel: at $30\%-40\%$ centrality. Down panel: at $60\%-92\%$
centrality. } \label{semi_peri}
\end{figure}

\section{III. Results}

In order to search for the origin of double-humped structure in
$\Delta\phi$ on the away-side, we use Monte Carlo to simulate the
process for an away-side jet going through the medium produced in
Au+Au collisions. There are three physics effects considered here:
$(1)$ with/without the flow effect; $(2)$ with/without the
variation of initial position for hard parton pair creation; $(3)$
with/without the initial momentum direction distribution of hard
parton. For simplicity, we classify our investigations into two
major categories: $(1)$ without flow effect; $(2)$ with flow
effect. For each category, there are four different cases
corresponding to different combinations of with/without hard
parton variation of creation point and initial moving angle
distribution.

First, let's consider four different cases with no flow effect. In
our model, we set $f_a=0$ to exclude the flow effect. For all the
four choices on the distributions of initial hard parton creation
position and the moving direction, we find that the away-side
di-hadron azimuthal angle distribution is a single peak on the
away-side for different parameter combinations. So we can conclude
that collective flow effect is essential for the double-humped
structure.

Second, the flow effect is included. When the hard parton is
assumed to be generated at a fixed point on the surface of the
medium system and its initial momentum is always directed at
$\theta=\pi$, no double-humped structure can be found on the
away-side di-hadron azimuthal distribution. The double-humped
structure does appear on the away-side when $(1)$ the hard parton
is generated at one point with distributed initial moving
direction; or $(2)$ the hard parton is generated on the
semi-circle surface of the collision system on the near-side
without diverse initial moving direction. But the simulated
$\Delta\phi$ correlation can't fit the experimental data well.
When the hard parton is generated on the semi-circle surface of
the medium with $\theta$ distribution for the initial moving
direction, we can fit the experimental data well with suitably
chosen parameters. In Fig. \ref{central} shows the di-hadron
correlation distribution from our model for the away-side together
with the experimental data \cite{star5}. The best fit corresponds
to $\omega=40, f_a=0.1, \lambda=0.05, \alpha=0.1$. A factor has
been multiplied to adjust the magnitude of the distribution. The
dip is around $\Delta \phi=\pi$ and double humps are at $\Delta
\phi\sim\pi\pm 1.1$.

The value of $\omega=40$ means that the mean magnitude of momentum
for the medium parton is $\frac{2}{\omega}=\frac{1}{20}$ that for
a hard parton. Considering the fact that the mean transverse
momentum of medium parton is about $\sqrt{2}T_c$, the value of
$\omega$ obtained in the fit corresponds to hard parton momentum
$k=20\times\sqrt{2}T_c\simeq 5\rm{GeV}$ for $T_c=0.17\rm{GeV}$.
This result seems reasonable for $P_{trig}\simeq 3-4 \rm {GeV}$.

In order to further check the model, we consider the dependence of
di-hadron azimuthal distribution on the centrality. We apply our
model with exactly the same parameters to the semi-central and
peripheral Au+Au collisions with the three physical effects
mentioned above. One can see from Fig. \ref{semi_peri} that the
corresponding experimental $\Delta \phi$ distributions can also be
reproduced very well. Also one can observe a transition from a
double-humped to single peak structure for the away-side azimuthal
distributions for Au+Au collisions from semi-central to
peripheral. This means that our model can describe the away-side
$\Delta \phi$ distributions for different colliding centralities.

\section{IV. Conclusion}
We have considered a model for the origin of the double-humped
structure on the away-side in heavy ion collisions. We simulated
the jet-medium interactions and obtained the di-hadron azimuthal
angle correlation. All of the parameters are fixed by fitting the
experimental data for central collisions. Then, we applied our
model to the semi-central and peripheral Au+Au collisions. The
corresponding azimuthal distribution on the away-side can also be
reproduced very well. From the results, we can conclude that the
transverse flow effect is the main origin of the double-humped
structure on the away-side in heavy ion collisions. Another
necessary effect for the double-humped structure is due to
distributions for the hard parton creation point and/or its
initial moving direction. Probably more realistic jet-medium
interactions need to be introduced to quantitatively explain the
experimental data.

\section{Acknowledgments}
This work was supported in part by the National Natural Science
Foundation of China under Grant Nos. 10635020 and 10775057, by the
Ministry of Education of China under Grant No. 306022 and project
IRT0624, and by the programme of Introducing Talents of Discipline
to Universities under No. B08033.

\end{document}